\documentclass[preprint,3p,12pt]{elsarticle}

\usepackage{amssymb}
\usepackage{xcolor}
\usepackage{hyperref}
\usepackage{amsmath}
\begin{document}

\begin{frontmatter}



\title{Basic Cycle Ratio: Cost-Effective Ranking of Influential Spreaders from Local and Global Perspectives}

\author[1]{Wenxin Zheng} %% Author name
\author[1]{Wenfeng Shi}
\author[1]{Tianlong Fan\corref{cor1}} 
\ead{tianlong.fan@ustc.edu.cn}
\author[1]{Linyuan Lü\corref{cor1}}
\ead{linyuan.lv@ustc.edu.cn}

\cortext[cor1]{Corresponding author}
\affiliation[1]{organization={School of Cyber Science and Technology, University of Science and Technology of China},%Department and Organization 
            city={Hefei},
            postcode={230026}, 
            country={China}}

%% Abstract
\begin{abstract}
Spreading processes are fundamental to complex networks. Identifying influential spreaders with dual local and global roles presents a crucial yet challenging task. To address this, our study proposes a novel method, the Basic Cycle Ratio (BCR), for assessing node importance. BCR leverages basic cycles and the cycle ratio to uniquely capture a node's local significance within its immediate neighborhood and its global role in maintaining network cohesion. We evaluated BCR on six diverse real-world social networks. Our method outperformed traditional centrality measures and other cycle-based approaches, proving more effective at selecting powerful spreaders and enhancing information diffusion. Besides, BCR offers a cost-effective and practical solution for social network applications.
\end{abstract}

%% Keywords

\end{frontmatter}



\section{Introduction}
Spreading is a pervasive dynamical process in complex networks \cite{newman2018networks, dorogovtsev2022nature, sha2025dynamic}, particularly evident in social contagion dynamics such as viral information diffusion \cite{ji2023signal,meng2025spreading} and behavioral contagion \cite{xie2022indirect, he2025impact}. Identifying important spreaders in social networks has become a significant research area that plays an important role in structure and functionality \cite{zhou2023identifying,chen2025dynamic}. 

Traditional centrality measures \cite{freeman2002centrality,korn2009lobby}, such as degree centrality \cite{lu2016vital} and betweenness centrality \cite{freeman1977bc}, have been instrumental in understanding the roles of individual nodes within a network. However, these measures often focus on either local or global network characteristics, providing a limited perspective on the true influence of a node \cite{curado2023novel}. The degree centrality, for instance, captures the immediate neighbors of a node but overlooks the broader network context, while global measures like betweenness centrality and PageRank \cite{brin1998anatomy} may not adequately reflect the local importance or the structural nuances of the network. 

To address these limitations, recent researches have ventured into developing measures that integrate both local and global structural information \cite{cao2024dynamic,nandi2025ic}, enhancing the accuracy of influential node identification. Among these, the cycle structure has been recognized for its significance in network connectivity and dynamics \cite{shi2013searching,sizemore2018cliques,lizier2012information}. Cycles-closed paths sharing identical start and end nodes create redundant connections in networks. This structural redundancy enhances robustness against disruptions and modulates the efficiency of information propagation. Nevertheless, a systematic understanding of the role that cycle structures play in shaping node influence remains limited. \cite{fan2021characterizing,shi2023cost,jiang2023searching,shi2025cycrank,zheng2025130830}. 

To address these gaps, we propose a novel method for ranking influential nodes based on the Basic Cycle Ratio (BCR).  The BCR is designed to quantify node importance on two levels: locally, by measuring a node's involvement within its basic cycle sets, and globally, by assessing its cohesive role in the overall network. This approach innovatively integrates two complementary perspectives: i) the number of basic cycles and ii) the cycle ratio. At the local level, BCR calculates the number of basic cycles a node participates in, reflecting its importance within immediate structures. Nodes with high local significance are crucial for network robustness because they create irreplaceable pathways for information flow. Globally, BCR evaluates a node's participation in cycles across the entire network to measure its cohesive role. This identifies pivotal nodes that bridge disparate network regions through their extensive cycle involvement.

To validate our approach's robustness, we tested the BCR on six real-world social networks against three classic centrality measures and two cycle-based benchmarks. The results show that BCR outperforms these benchmarks by identifying spreaders that achieve a superior spreading effect. Moreover, our method is cost-effective and provides more solutions in practice. 

The rest of the paper is organized as follows. Section \ref{preliminaries} introduce preliminaries. The proposed method BCR is presented in Section \ref{method}. Additionally, a selection of real-world networks and indicators is employed to illustrate the efficiency and robustness of the proposed method in Section \ref{results}. Finally, the conclusion is presented in Section \ref{conclusion}.

\section{Preliminaries}\label{preliminaries}
\subsection{Centrality measures}
Degree Centrality (DC) \cite{lu2016vital}. In a network $G(V,E)$, the adjacency matrix $A=\{a_{ij}\}$ is defined as $a_{ij}=1$ when there is an edge between node $i$ and $j$, and $a_{ij}=0$ otherwise. The degree centrality of each node in the network is given by $\mathrm{DC}_i=\sum_{j=1}^Na_{ij}$.  Degree centrality measures the importance of a node from a local perspective. 

Coreness \cite{kitsak2010identification}. Coreness is a measure that assesses the degree to which a node is part of a densely connected core within a network. It assigns a core number based on the highest k-core to which the node belongs, where the k-core is a subnetwork obtained through k-shell decomposition and nodes within it have at least k connections. The coreness is an approach that reveals the global network structure through the iterative application of local information.

Betweenness centrality (BC) \cite{freeman1977bc}. A node is considered significant and possesses high betweenness centrality if it lies on the sole path that other nodes must traverse. The BC score of a node is defined as
\begin{equation}
    \mathrm{BC}_i=\sum_{i\neq s,i\neq t,s\neq t}\frac{g_{st}^i}{g_{st}},
\end{equation}
where $g_{st}$ represents the number of shortest paths between nodes $s$ and $t$, $g^i_{st}$ denotes the number of those shortest paths that pass through node $i$.  Betweenness centrality measures the importance of a node from a global perspective.

\subsection{Number of basic cycles (NC)}
Shi et al. \cite{shi2023cost} proposed an indicator NC based on the cycle number, which is derived from the basic cycles. First, the basic cycle set $B$ of network $G(V, E)$ is calculated through the spanning tree $T$ of the network, thus basic cycles $c$ and their set $B$ are defined as:
\begin{equation} 
c_k = {(s, t) \cup P_{st}}, 
\end{equation}
\begin{equation}
    B=\{c_1,c_2,\ldots,c_k\},
\end{equation}
where $(s, t)$ is an edge satisfying $(s, t) \in E$ and $(s, t) \notin T$. The path $P_{st}$ is the unique path in $T$ linking node $s$ to $t$. Basic cycles provide a robust and scalable foundation for analyzing key node interactions and assessing their influence in complex networks.

Subsequently, the number of basic cycles passing through nodes provides a metric where the importance of a node $i$ is given by
\begin{equation}
    \mathrm{NC}_i=\sum_{c\in B}\delta(c,i),
\end{equation}
where $\delta(c, i)$ is an indicator function, $\delta(c, i)=1$ if node $i$ is in cycle $c$, and $\delta(c, i)=0$ otherwise. The authors showed that NC can identify multiple influential spreaders with outstanding propagation performance and low initial cost. 

\subsection{Cycle ratio (CR)}
Fan et al. \cite{fan2021characterizing} defined the shortest cycles of node $i$ as the cycles containing node $i$ with the smallest size, denoting their set as $S_i$. Based on this concept, they introduced  a cycle number matrix $C$, where the element $c_{ij}$ is defined as follows: 
\begin{equation}
    c_{ij}=\begin{cases}\text{the number of cycles in}\ S_i\ \text{that pass through nodes}\ i\ \text{and}\ j,\ i\neq j,\\\text{the number of cycles in}\ S_i,\ i=j.\end{cases}
\end{equation}
Subsequently, they introduced the concept of cycle ratio (CR) to quantify the degree to which node $i$ is involved in the shortest cycles of other nodes. The CR is calculated using the following formula: 
\begin{equation}
    \mathrm{CR}_i=\left\{\begin{matrix}0,&c_{ii}=0,\\\sum_{j,c_{ij}>0}\frac{c_{ij}}{c_{jj}},&c_{ii}>0.\end{matrix}\right.
\end{equation}
And they showed that the CR outperforms DC, HI, and KC in the early stages of spreading.

\section{Method}\label{method}
We propose a new indicator combining two concepts: i) the number of basic cycles to capture the information of the topology of a network, and ii) the cycle ratio to detect the real strength of each node in a basic cycle set.

\subsection{Formal analysis}\label{formal}
Recently, basic cycles demonstrated the ability to reveal the deep structural characteristics of networks \cite{shi2023cost}. Meanwhile,  in \cite{fan2021characterizing}, the combination of the shortest cycles and the cycle ratio efficiently captured nodes that strengthen the network, as a significant amount of information is transmitted through them. 

Our proposed indicator captures both the network's structural properties and the influence of individual nodes on information flow. The methodology can be calculated in three steps.

\subsubsection{Step 1: Calculation of the basic cycles in networks}
We calculate the basic cycle set of the network, denoted by $B=\{c_1,c_2,\ldots,c_k\}$, where $k$ is the total number of distinct basic cycles. 

\subsubsection{Step 2: Calculation of the basic cycle number matrix}
While the cycle ratio (CR) derives its values from a cycle number matrix, its reliance on shortest cycles limits its ability to capture the network’s topological structure. To address this limitation, we propose constructing the cycle number matrix using basic cycles.

We initialize the basic cycle number matrix $C$ for a given network $G(V,E)$ with dimension $N * N$. As shown in Eq.\eqref{C}, each element of matrix $C$ represents the number of basic cycles in the network that contain both node $i$ and $j$.
\begin{equation}
    C=\begin{bmatrix}c_{11}&c_{12}&\cdots&c_{1N}\\c_{21}&c_{22}&\cdots&c_{2N}\\\vdots&\vdots&\ddots&\vdots\\c_{N1}&c_{N2}&\cdots&c_{NN}\end{bmatrix}.
    \label{C}
\end{equation}
The element $c_{ij}$ in the cycle number matrix $C$ is as follows:
\begin{equation}
    c_{ij}=\left\{\begin{matrix}c_{ii}=\sum_{c\in B}\delta(v_i\in c),i=j,\\\sum_{c\in B}\left(\delta(v_i\in c)\cdot\delta(v_j\in c)\right),i\neq j,\end{matrix}\right.
\end{equation}
where $\delta$ is an indicator function, if the condition in parentheses is true, then $\delta=1$, otherwise $\delta=0$.

\subsubsection{Step 3: Calculation of the basic cycle ratio of each node}
Based on the basic cycle number matrix $C$, we can calculate the importance of each node as follows:
\begin{equation}
\mathrm{BCR}_i=\begin{cases}0,c_{ii}=0,\\\sum_{j,c_{ij}>0}\frac{c_{ij}}{c_{jj}},c_{ii}>0,\end{cases}  
\end{equation}
where $c_{ii}$ denotes the number of basic cycles that node $i$ participates in, while $c_{ij}$ represents the number of basic cycles in which node $i$ and $j$ are involved.

\subsection{Calculation of the basic cycle ratio: a sample}
\begin{figure}[t]
\centering
\includegraphics[width=1.0\textwidth]{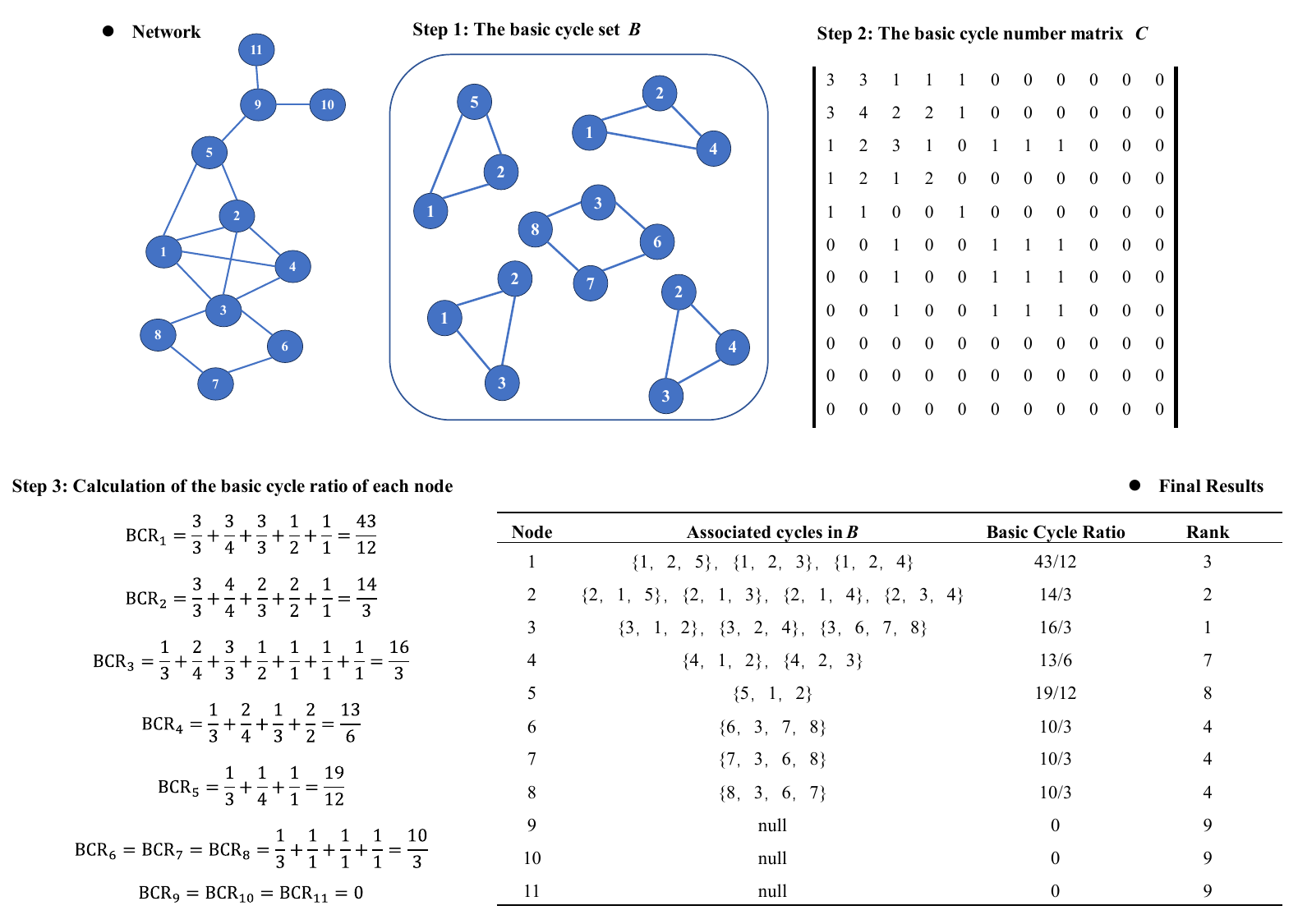}
\caption{Basic cycle ratios of nodes in an example network.}
\label{fig:sample}
\end{figure}

To clarify the methodology used to calculate the importance of nodes through the basic cycle ratio, a sample is presented. In Figure \ref{fig:sample}, we show a network with 11 nodes and 15 edges. We calculate the importance values following the three steps described in Section \ref{formal}, and give the node ranks as final results. The results illustrate that node 3 has the highest BCR values, making it the most important node, followed by nodes 2 and 1. Nodes 9, 10, and 11 is the least important nodes as they do not participate in any cycle, resulting in a BCR value of zero.

\section{Result}\label{results}
This section validates the effectiveness of BCR by comparing it with local (Degree), glocal (Coreness), global (Betweenness) centralities, and the recent measures (CR and NC) based on cycle structure. We evaluate the proposed measures on six real-world social networks: Collaboration, Email, Ia-facebook, Soc-epinions, Soc-facebook, and Soc-hamsterster. These networks exhibit diverse sizes and structural characteristics.

\subsection{Dataset}
The introduction of real social networks is as follows:

Collaboration \cite{newman2001structure}: This network contains a collaborative network of scientists who have published preprints of high-energy theories.

Email \cite{guimera2003self}: This network depicts the flow of emails among employees at Rovira i Virgili University in Spain. 

Ia-facebook \cite{rossi2015network}: This network is similar to Facebook, and it includes users who have either sent or received at least one message. 

Soc-epinions \cite{richardson2003trust}: This network represents an online social platform of Epinions.com, where users engage in a trust-based relationship.  

Soc-facebook \cite{traud2012social}: This network, extracted from Facebook, consists of people with edges representing friendship ties.

Soc-hamsterster \cite{rossi2015network}: This network illustrates the Hamsterster social platform, where nodes symbolize users and edges signify connections of friendship or kinship. 

The structural information of these networks is detailed in Table \ref{tab:networks}. Here, \(N\) and \(E\) are the number of nodes and edges in the network respectively. \(D\) signifies the density of these networks, whereas \(C\) indicates the average clustering coefficient. ⟨\textit{k}⟩ denotes the average degree of the nodes. 

\begin{table}[t]
    \centering
    \caption{ The detailed structural information of the real networks}
    \begin{tabular}{cccccc}
    \hline
        Network & \(N\)&  \(E\)& \(D\)& \(C\)& ⟨\textit{k}⟩\\ \hline
        Collaboration & 5835 & 13815 & 0.000812 & 0.506193 & 4.7352  \\ 
        Email & 1133 & 5451 & 0.008500 & 0.220176 & 9.6222\\ 
        Ia-facebook & 1266 & 6451 & 0.008056 & 0.068350 & 10.1912  \\ 
        Soc-epinions & 3000 & 48922 & 0.010875 & 0.184972 & 32.6147  \\ 
        Soc-facebook & 1510 & 32984 & 0.028951 & 0.316606 & 43.6874  \\ 
        Soc-hamsterster & 2000 & 16097 & 0.008053 & 0.539978 & 16.0970\\ \hline
    \end{tabular}
    \label{tab:networks}
\end{table}

\subsection{Correlation between BCR and benchmark indicators}
Before discussing the ability of BCR, we analyze its correlation with other benchmarks. We use Kendall's tau (\(\tau\)) \cite{kendall1938new} to measure the correlation computationally and utilize key node visualization to elucidate the interrelationships visually. 

Kendall’s \(\tau\) correlation coefficient serves as a metric for quantifying the similarity between two ranking lists. It spans a range from -1 to 1, where higher values signify greater similarity and lower values denote greater dissimilarity. Kendall’s tau is defined as:
\begin{equation}
    \tau_b=\frac{2(N_c-N_d)}{N(N-1)},
\end{equation}
where $N_c$ is the number of concordant pairs, and $N_d$ is the number of discordant pairs in a two-by-two comparison. 

\begin{figure}[t]
\centering
\includegraphics[width=0.8\textwidth]{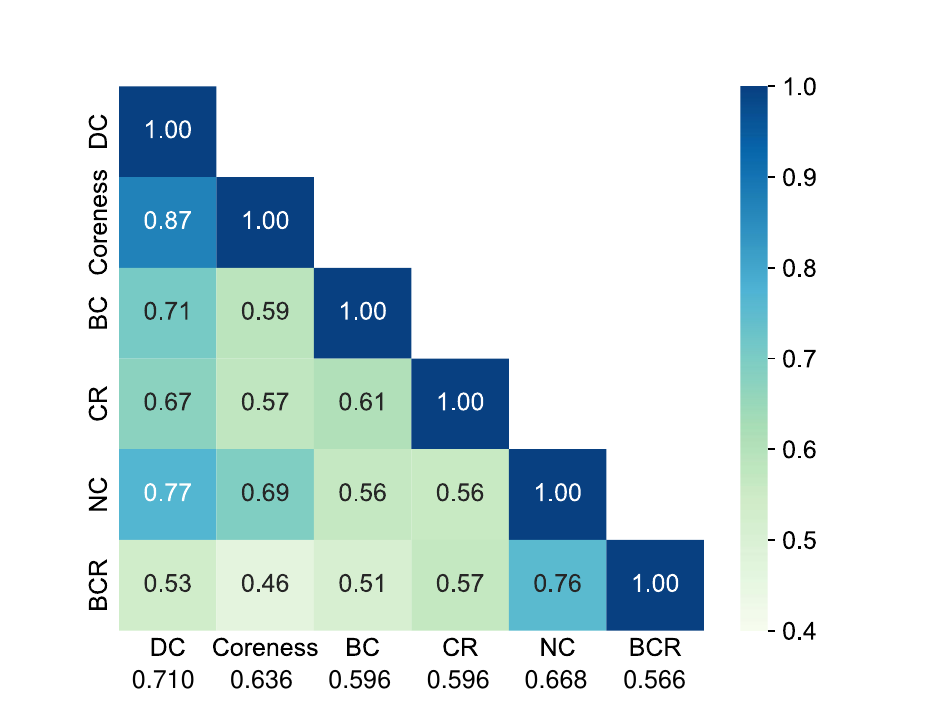}
\caption{Average Kendall’s tau (\(\tau\)) among the six indicators over six real-world networks. The values in each cell represent the average correlation between a pair of indicators, and the color intensity corresponds to the magnitude of \(\tau\). }
\label{fig:kendall}
\end{figure}

As shown in Figure \ref{fig:kendall}, the correlations among cycle-based methods (CR and NC) and our proposed indicator BCR are relatively high, owing to their common structural focus. In contrast,  the correlation between BCR and classical centrality methods (DC, Corness, and BC) is relatively low, as they are cycle-based, degree-based, and path-based, respectively. Notably, BCR exhibits the lowest average correlation (0.566) with all other benchmarks. This distinctiveness suggests that the node rankings provided by BCR capture unique information, offering a perspective not available from the existing measures. 

\begin{figure}[t]
\centering
\includegraphics[width=0.8\textwidth]{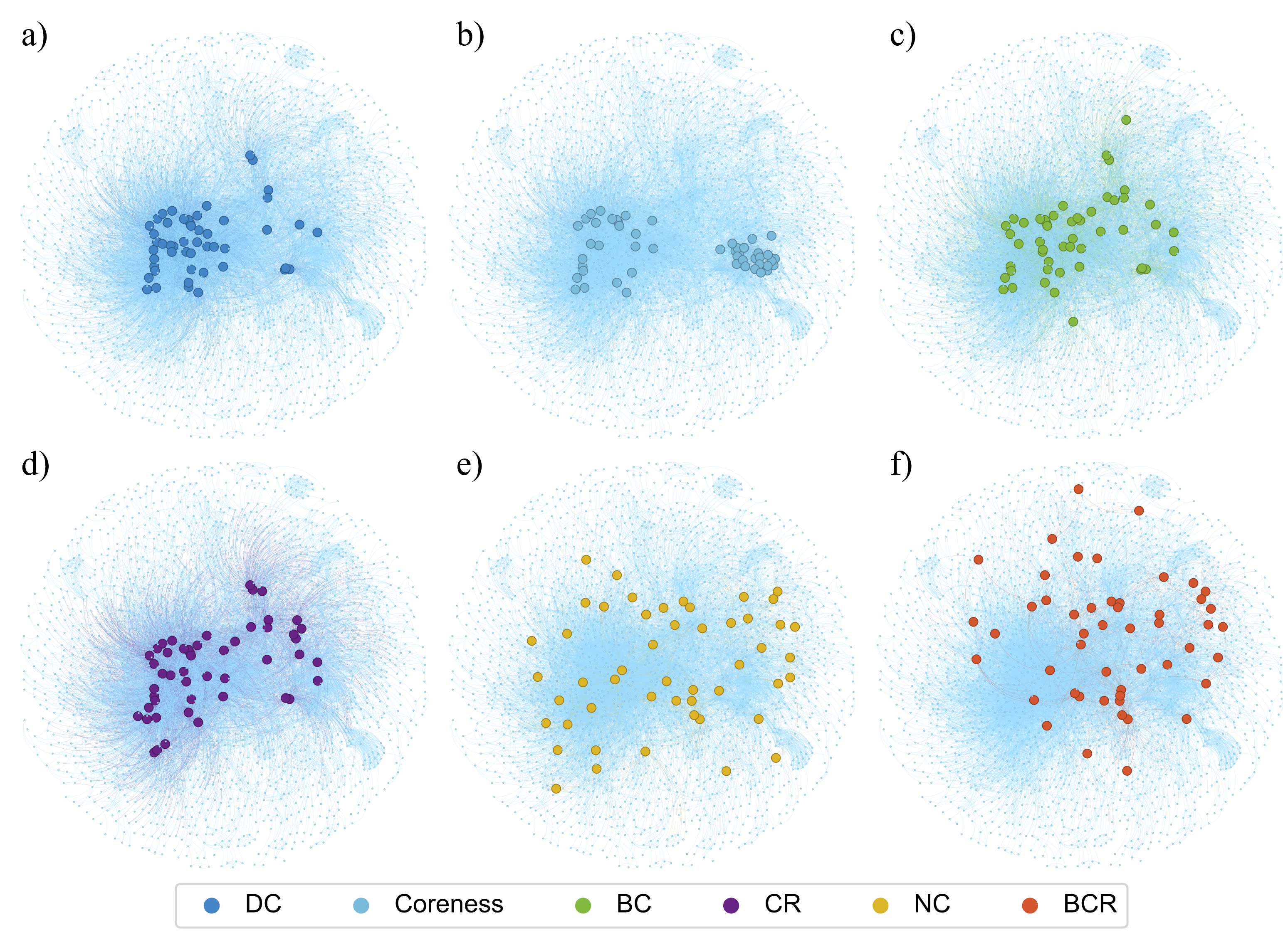}
\caption{Visualization of the top-50 ranked nodes identified by six indicators in the soc-hamsterster network. Across all plots, node importance is mapped to color (see legend) and size (larger indicates greater importance), while a fixed layout is maintained for consistent node positioning. (a) DC; (b) Coreness; (c) BC; (d) CR; (e) NC; (f) BCR.}
\label{fig:visulization}
\end{figure}

Figure \ref{fig:visulization} presents a visualization of the top-50 important nodes selected by each indicator in the soc-hamsterster network. Intuitively, important nodes selected by DC and Coreness are closely connected and clustered in a certain area, which is consistent with the so-called rich club phenomenon \cite{zhou2004rich,colizza2006detecting}. In contrast, the important nodes identified by BCR are more widely distributed. Unlike BC and CR, the nodes identified by BCR are more evenly distributed and better connected to marginal communities. While BCR shares with NC the ability to identify key nodes within dense communities in a balanced manner, it outperforms NC by also selecting nodes that bridge multiple communities. This demonstrates that BCR considers not only important nodes in dense communities but also inter-community nodes, covering the margin of the network.

\subsection{Individuation of node rankings}
A key aspect of evaluating a ranking indicator is its ability to resolve ambiguities. To effectively distinguish the importance of nodes, a desirable property is the assignment of a unique score to each node, thereby producing a clear, unambiguous ranking. Accordingly, we use a measure  \(\gamma(\cdot)\) \cite{wen2020vital} to detect the ability of each indicator to assign unique scores to nodes as follows:
\begin{equation}
    \gamma(\cdot)=\frac{N_{S}(\cdot)}{|N|},
\end{equation}
where \(N_{S}(\cdot)\) is the number of nodes with a unique score assigned by one method, and \(|N|\) is the number of nodes in the entire network, \(\gamma(\cdot)\) is the individuation of the method. The core idea of this method is that the higher the individuation of an indicator, the more effective it is perceived to be.

\begin{figure}[t]
\centering
\includegraphics[width=0.8\textwidth]{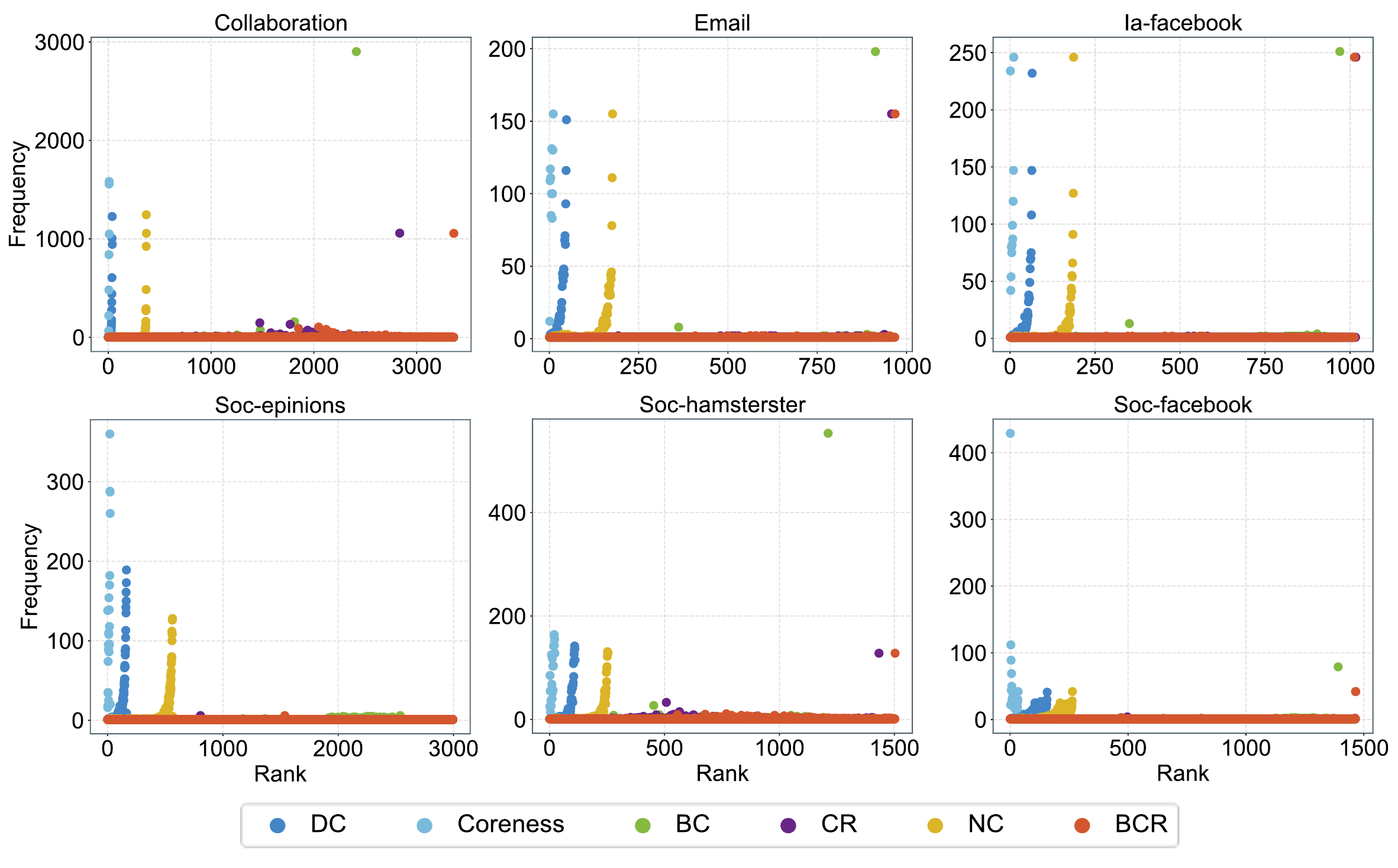}
\caption{The frequency of nodes of each class for different indicators across six real networks. The $x$-axis shows the top node ranks, and the $y$-axis indicates the frequency of shared scores.}
\label{fig:individuation}
\end{figure}

We evaluated the ability of each indicator to assign unique scores to nodes across six real-world networks. Figure \ref{fig:individuation} displays the frequency distribution of node rankings for the six indicators. BCR stands out with a significant advantage, as it assigns a unique score to nearly every node in most networks. In contrast, DC, coreness, and NC perform poorly, as a large number of nodes share identical ranks. This issue is most acute for coreness, where the majority of nodes share the same rank in the initial phase, severely limiting its ability to differentiate node importance. Although BC and CR also exhibit a strong capacity for distinction, they are less robust than BCR. As detailed in Table \ref{tab:individuation} (where the highest \(\gamma(\cdot)\) values are in bold), BCR achieves the best performance in five out of the six networks. 

\begin{table}[t]
    \centering
    \caption{The individuation \(\gamma(\cdot)\) of different methods in complex networks.}
    \begin{tabular}{ccccccc}
    \hline
        Network & DC & Corness & BC & CR & NC & BCR  \\ \hline
        Collaboration & 0.0067 & 0.0017 & 0.4135 & 0.4859 & 0.0636 & \textbf{0.5760 } \\ 
        Email & 0.0424 & 0.0097 & 0.8058 & 0.8455 & 0.1562 & \textbf{0.8544}  \\ 
        Ia-Facebook & 0.0513 & 0.0087 & 0.7662 & \textbf{0.8033} & 0.1477 & 0.8002  \\ 
        Soc-epinions & 0.0553 & 0.0083 & 0.8567 & 0.9980 & 0.1883 & \textbf{0.9983}  \\ 
        Soc-facebook & 0.1040 & 0.0225 & 0.9212 & 0.9701 & 0.1748 & \textbf{0.9702}  \\ 
        Soc-hamsterster & 0.0555 & 0.0115 & 0.6060& 0.7165 & 0.1270 & \textbf{0.7515}  \\ \hline
    \end{tabular}
    \label{tab:individuation}
\end{table}

\subsection{Spreading performance of BCR }
In this section, we employ the susceptible-infectious-recovered (SIR) model \cite{anderson1991infectious} to evaluate the performance of BCR in identifying influential spreaders. In this model, each node is in one of three states: susceptible, infectious, or recovered. The dynamics are defined by two probabilities: an infected node infects each susceptible neighbor with probability \(\beta\), and an infected node recovers with probability \(\mu\). Recovered nodes gain permanent immunity. The top-ranked nodes selected by the indicators are set as the initial infected seeds, while all others are initially susceptible. The parameters are set to $\mu=0.5$ and $\beta=\beta_c=\langle k\rangle/(\langle k^2\rangle-2\langle k\rangle)$, where \(\langle k\rangle\) and \(\langle k^2\rangle\) are the network's mean degree and mean squared degree, respectively.  

\begin{figure}[t]
\centering
\includegraphics[width=0.8\textwidth]{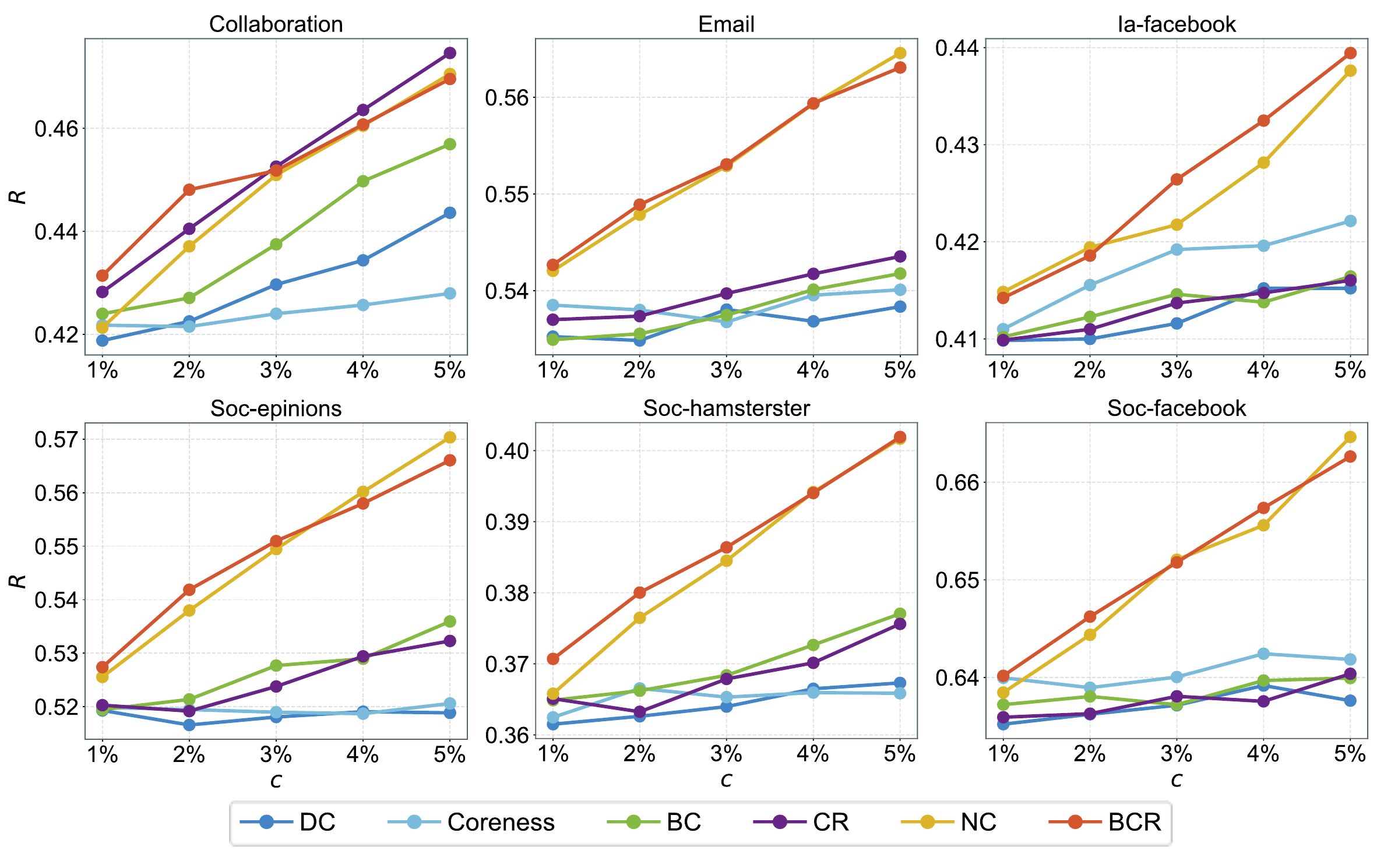}
\caption{Spreading ability of each indicator at \(\beta\)= 1.5\(\beta_c\) for different sizes of the initial seed set $c$ (from the top 1\% to 5\%) over six empirical networks. The $x$-axis denotes the initial seed proportion ($c$), and the $y$-axis indicates the proportion of infected nodes ($R$) in the spreading.}
\label{fig:SIR_node}
\end{figure}

To evaluate robustness against parameter settings, we first fixed the infection rate at $\beta=1.5\beta_c$ and varied the size of the initial seed set $c$ from the top 1\% to 5\% of nodes. The resulting spreading ability (\(R\) )  for each indicator is shown in Figure \ref{fig:SIR_node}. BCR, indicated by the red line, consistently ranks highly, outperforming benchmark indicators and frequently attaining the best performance. 

\begin{figure}[t]
\centering
\includegraphics[width=0.9\textwidth]{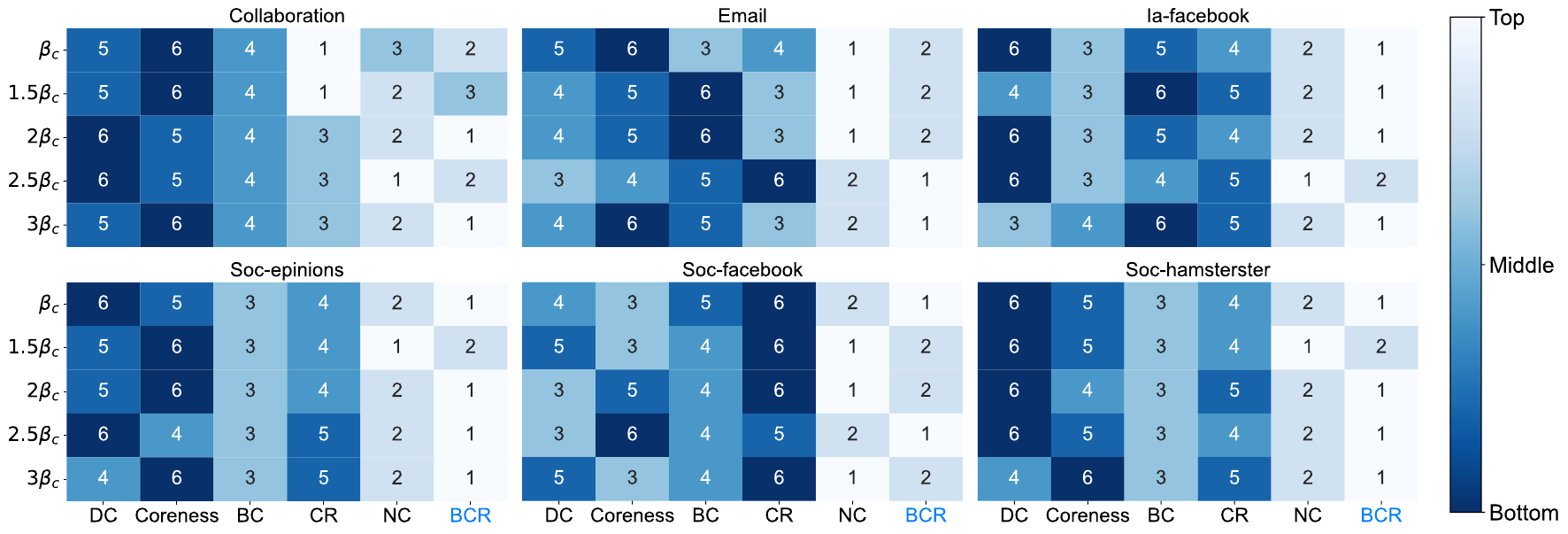}
\caption{Spreading ability of each indicator among the top-3\% spreaders for different infection probabilities across six empirical networks. The color indicates the ranking of the final contagion range, with brighter colors indicating a greater contagion scope.}
\label{fig:SIR_beta}
\end{figure}

We further assessed the indicators by adjusting the infection rate across a range of values ($\beta=\beta_c$, $1.5\beta_c$, $2\beta_c$, $2.5\beta_c$, $3\beta_c$) while fixing the initial seed set to the top  3\% of nodes. For clarity, Figure \ref{fig:SIR_beta} visualizes the ranking of the final outbreak size under each method. The results confirm that BCR consistently enables wider dissemination than other methods across all tested infection rates. 

\begin{figure}[t]
\centering
\includegraphics[width=0.8\textwidth]{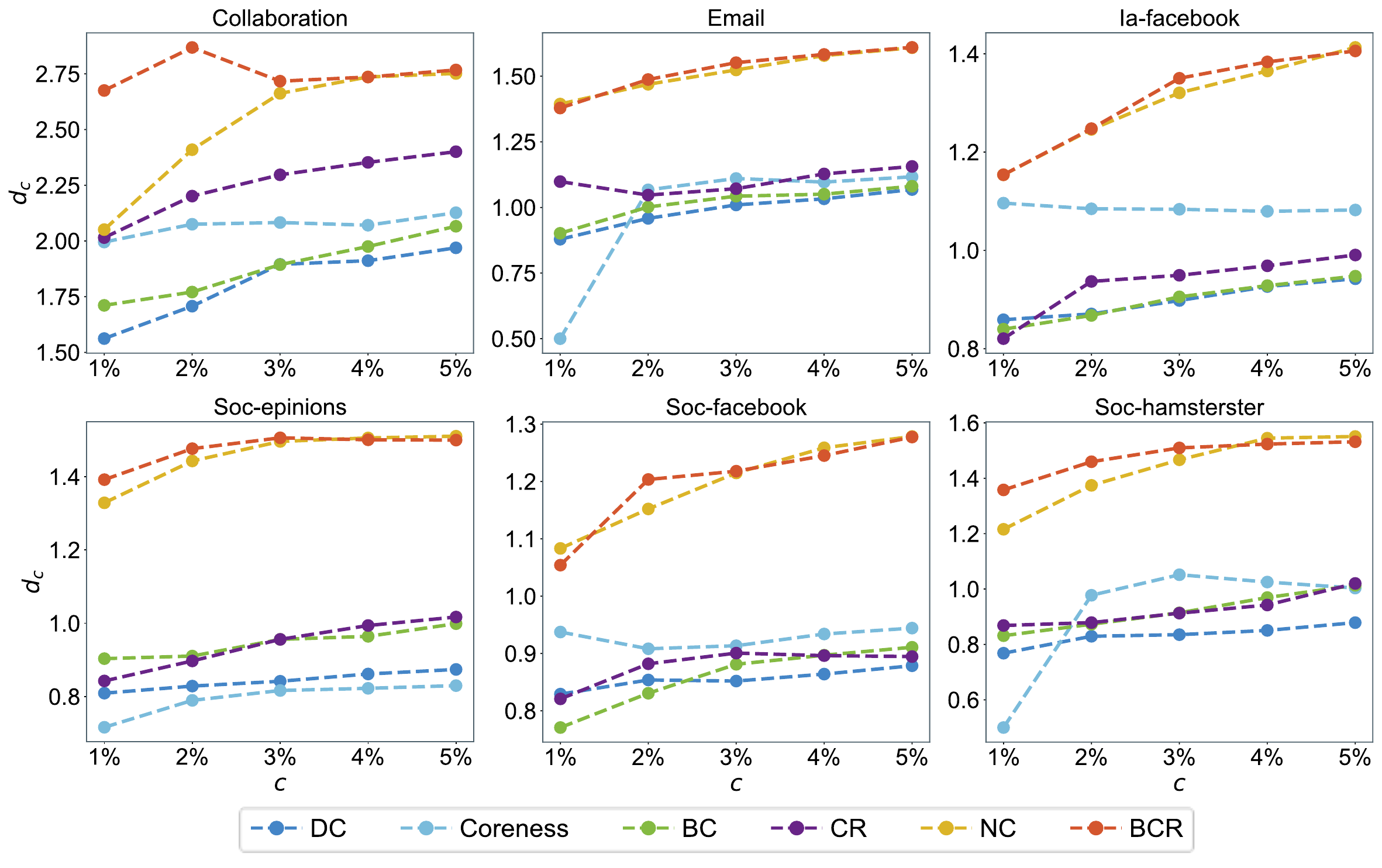}
\caption{Comparison of the average shortest distance $d_c$ among the node groups $c$ ($c=1\%, 2\%, 3\%, 4\%$, and $5\%$) selected by six indicators in the six empirical networks. The $x$-axis represents the seed set proportion $c$, and the $y$-axis indicates the corresponding average distance $d_c$,  where a larger value signifies greater spatial dispersion among the key nodes. }
\label{fig:distance}
\end{figure}

We further discuss the original advantages of BCR. When multiple spreaders are considered simultaneously, the distance between them is a key parameter determining the spreading extent \cite{influence_distance}.  From this aspect, we analyze the average shortest distance ($d_c$) among node groups of size  $c$ ( $c=1\%, 2\%, 3\%, 4\%$, and $5\%$) identified by six indicators. As shown in Figure \ref{fig:distance}, the average distance of the node groups identified by BCR is the largest across the six real-world networks. This indicates that nodes with high BCR scores are typically distant from one another. Therefore, BCR reduces the overlap of the areas influenced by different spreaders, resulting in its excellent spreading performance. 

\subsection{Initializing cost of BCR}
In the above section, we learned BCR's spreading capability. And then we will focus on BCR's advantages in identifying important node groups. A notable advantage of BCR lies in its low-cost identification of influential spreaders.

In real life, information spreading often entails initial costs. For instance, influential bloggers or celebrities often require financial compensation for posting or delivering advertisements. Ji et al. \cite{ji2017effective} examine the initializing cost of selecting spreaders by measuring his/her impact as well as scarcity, which can be defined as follows:
\begin{equation}
    \lambda=\sum_{i=1}^{c}\frac{k_{i}}{p(k_{i})},
\end{equation}
where \(c\) is the top $c$ selected spreaders, degree \(k_i\) represents the node's impact, and probability \(p(k_i)\) shows its scarcity. 

\begin{figure}[t]
\centering
\includegraphics[width=0.8\textwidth]{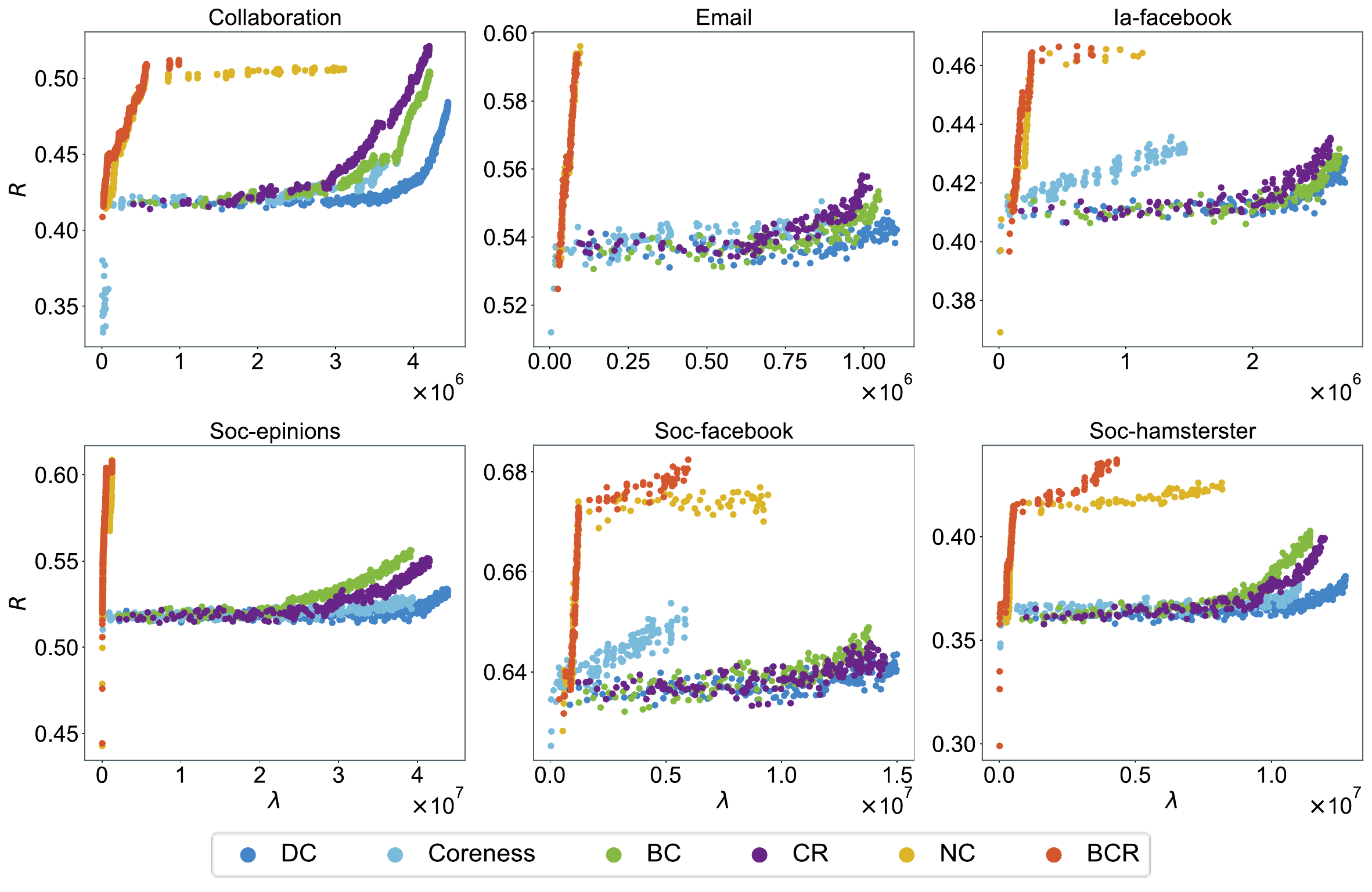}
\caption{Overall spreading ability (\(R\)) and cost (\(\lambda\)) of the six networks. The $x$-axis represents the total cost of the top-\(c\) nodes, and the $y$-axis represents their corresponding spreading ability.}
\label{fig:cost}
\end{figure}

Figure \ref{fig:cost} evaluates the trade-off between spreading ability (\(R\)) and the cost (\(\lambda\)) of top-ranked spreaders, with the source spreader fraction \(c\) spans from 2 to 10\% of the network size. The results clearly show that BCR is the most cost-effective, achieving a higher \(R\) at an equivalent \(\lambda\) than all benchmark methods. Furthermore, BCR reaches this superior performance with a lower overall cost, as indicated by its smaller maximum \(\lambda\) values. Overall, this dual optimization of spreading efficacy and resource allocation provides strategic advantages for information propagation system design. 

\subsection{Multiple solutions of BCR}
The selection of basic cycles depends on the network's spanning tree, which may vary across different realizations. This variation could lead to fluctuations in the top spreaders identified by NC and BCR. To demonstrate the robustness of BCR against this randomness, we conducted 30 independent realizations by generating different spanning trees. For each realization, we computed the overall spreading ability (\(R\)) of the top-\(2\%\) spreaders. Table \ref{tab:multiple} presents the average $R$ values over these 30 realizations, with the highest value highlighted in bold. 

The results show that the average \(R\) of BCR remains dominant across all indicators on 4 out of 6 networks, confirming its robustness against the randomness in spanning tree selection. Furthermore, the minimal variance across the 30 realizations indicates that BCR's performance is highly stable and largely unaffected by this variation. This stability implies that BCR provides more options for solutions when some nodes are unsatisfactory and ensures that the spreading performance is at a high level.

\begin{table}[t]
\footnotesize
    \centering
    \caption{Overall spreading ability (\(R\)) of the top-\(2\%\) nodes of six indicators on the six networks. For BCR and NC, values represent the mean $R$ calculated over 30 realizations (variance indicated in parentheses), each based on a different spanning tree. }
    \begin{tabular}{ccccccc}
    \hline
        Network & DC & Corness & BC & CR & NC & BCR  \\ \hline
        Collaboration & 0.422512 & 0.421505 & 0.427070& \textbf{0.440493} & 0.437937(7.65E-06) & 0.438381(7.37E-06)  \\ 
        Email & 0.534848 & 0.538008 & 0.535543 & 0.537367 & 0.545880(3.18E-06)& \textbf{0.546604}(5.80E-06)  \\ 
        Ia-Facebook & 0.409997 & 0.415545 & 0.412267 & 0.410993 & 0.424244(2.66E-06) & \textbf{0.424679}(2.42E-06)  \\ 
        Soc-epinions & 0.516591 & 0.519467 & 0.521368 & 0.519177 & 0.535336(3.88E-06) &\textbf{0.535493}(4.46E-06)  \\ 
        Soc-facebook & 0.636230& 0.638954 & 0.638055 & 0.636291 & \textbf{0.647670}(1.20E-06)& 0.647518(1.31E-06)  \\ 
        Soc-hamsterster & 0.362625 & 0.366547 & 0.366217 & 0.363258 & 0.376063(3.85E-06) & \textbf{0.377590}(3.87E-06)\\ \hline
    \end{tabular}
    \label{tab:multiple}
\end{table}

\section{Conclusion}\label{conclusion}
Current studies on influential node identification advocate a holistic framework that integrates topological information with the strength of each node in the structure. In response to this paradigm, we propose the basic cycle ratio (BCR), a novel metric that synergistically combines the concept of basic cycles with the cycle ratio. Specifically, BCR first utilizes basic cycles to encode the local network topology; it then employs the cycle ratio to quantify the real strength of each node within the cycle set. 

The effectiveness of BCR is rigorously evaluated against classical centralities and cycle-based measures. Experimental results demonstrate that BCR achieves superior spreading efficiency, maintains cost-effectiveness, and supports flexible multi-spreader selection. It exhibits higher discriminative power and robustness across diverse networks, providing richer node ranking information and broader dissemination coverage than benchmarks, while ensuring stable performance under varying conditions. This dual optimization of efficacy and resource efficiency makes BCR a strategically advantageous solution for practical information propagation systems. 

Nonetheless, this work has two limitations. First, the proposed BCR method is currently designed for undirected and unweighted networks, which restricts its applicability to more complex networks involving direction and edge weights.  Future work could extend BCR by incorporating additional features of cycles—such as direction, weight, and length—to enhance its adaptability to a wider range of network types. Second, BCR's fundamental reliance on cycle structures prevents its application to tree-like or acyclic networks. A promising direction would be to integrate cycle analysis with other topological features, enabling the method to leverage structural information even in networks with very few or no cycles. 

\section*{CRediT authorship contribution statement}
Wenxin Zheng: Conceptualization, Methodology, Software, Writing – original draft. Wenfeng Shi: Conceptualization, Methodology, Validation, Writing - review \& editing. Tianlong Fan:  Conceptualization, Methodology, Validation, Writing - review \& editing, Supervision, Funding acquisition.  Linyuan Lü: Conceptualization, Writing – review \& editing, Supervision, Funding acquisition. 

\section*{Data and code availability}
Data and code are available at \href{https://github.com/Wenxin02/BasicCycleRatio}{https://github.com/Wenxin02/BasicCycleRatio} .

\section*{Competing interests}
The authors declare no competing interests.

\section*{Acknowledgments}
This work was supported by the National Natural Science Foundation of China (Grant Nos. T2293771, 62503447), the STI 2030 Major Projects (Grant No. 2022ZD0211400), the China Postdoctoral Science Foundation (Grant No. 2024M763131), the Postdoctoral Fellowship Program of CPSF (Grant No. GZC20241653), and the New Cornerstone Science Foundation through the XPLORER PRIZE.

% \bibliographystyle{elsarticle-numClean} 
% \bibliography{ref}

\end{document}